% Tex version of 1978 preprint that was never published
\input jnl
\def\sq#1{\lower1.5ex\hbox{$\sim$}\mkern-14mu #1}

{\singlespace
\line{\hfil CALT-68-709}
\line{\hfil DoE RESEARCH AND}
\line{\hfil DEVELOPMENT REPORT}
}
\vskip .5cm
\centerline { The Family Group in Grand Unified Theories\footnote*{Work 
supported in part by the U.S. Department of Energy under Contract No. 
EY76-C-03-0068.}}
\centerline {P. Ramond\footnote{**}{Robert Andrews Millikan Senior 
Research Fellow.}}
\centerline {\underbar{California Institute of Technology, Pasadena, California 91125}}
\bigskip
\centerline {Invited talk at the Sanibel Symposia}
\centerline {February 1979}
\bigskip
\centerline {ABSTRACT}
We review the known ways of incorporating and breaking symmetries in a 
renormalizable way.  We summarize the various grand unified theories based 
on $SU_5,~SO_{10}$, and $E_6$ as family enlargement groups.  An $SU_5$ 
model with an $SU_2$ gauged family group is presented as an illustration.  
In it, the e-family (i.e., e,u and d) is classically massless and acquires 
calculable mass corrections.  The family group is broken by the same agent 
that does the superstrong breaking.  Finally, we sketch a way of unifying 
the family group with $SU_5$ into $SU_8$.
\vfill\eject

The most important task facing model builders is to find the symmetries of 
the action that describes the fundamental interactions.  With the 
realization that strong, weak and electromagnetic forces are most probably 
described by Yang-Mills theories came the possibility that all these 
interactions, although very different at our scale, are but manifestations 
of the same master Yang-Mills theory$^{1,2)}$.  In this talk we shall 
assume:  1) the validity of QCD$^{3)}$, the gauge theory based on color 
$SU_3$ to describe strong interactions, 2) the 
Glashow-Weinberg-Salam-Ward$^{4)}~SU_2 \times U_1$ Yang-Mills model as the 
correct theory for the weak and electromagnetic interactions, 3) that 
gravity can be neglected in first approximation as the scales of interest 
will still be several orders of magnitude away from those where quantum 
gravity is thought to be important.  This survey will be divided in three 
parts:

I) A general theoretical discussion of the known allowed ways to 
build renormalizable theories with all types of symmetries, broken 
explicitly, spontaneously, or not at all.

II) A description of the zoo of elementary fermions and of the 
unification picture it suggests.  The concept of families of elementary 
particles will be seen to emerge.  The $SU_5^{(2)}$ model will be reviewed 
as well as the most likely family enlargement models based on 
$SO_{10}~^{5,6)}$ and $E_6~^{7)}$.

III) Last, but not least, a presentation of models that postulate 
interactions between the families ruled by ``family groups''.  An 
illustrative model of this type will be discussed where the $e,~u,$ and 
$d$ masses are zero classically, but acquire calculable quantum 
corrections.

\noindent { I)  \underbar{Allowed symmetries of renormalizable Lagrangians and 
their breaking.}}

It is convenient to split up the Lagrangian into several parts
$${\cal L} = {\cal L}_{\rm {kin}} + {\cal L}_{\rm{int}}^{\rm g} + {\cal 
L}_{\rm{int}}^Y + {\cal L}_{\rm{int}}^{\rm sc}\ ,$$
where ${\cal L}_{\rm{kin}}$ contains the kinetic terms of the particles 
which we take to have spins 0, 1/2, 1.  The spin-1 kinetic part will 
always be of the Yang-Mills variety and thus will include 
self-interactions.  ${\cal L}_{\rm{int}}^{\rm g}$ describes the interactions of 
the spin-1 gauge fields with the spin-0,-1/2 fields, and vanishes as the 
gauge coupling, g, tends to zero.  ${\cal L}_{\rm{int}}^Y$ contains 
fermion mass terms (if any) and fermion-fermion-spinles boson 
interactions.  Finally ${\cal L}_{\rm{int}}^{\rm sc}$ displays the 
self-interactions of the spinless bosons, including their mass terms, and 
is equal to minus the classical potential.  Before discussing each term in 
detail, let us ask what possible symmetries ${\cal L}$ can have.  These 
come in two categories - continuous and discrete.  Continuous symmetries 
can be of the graded-Lie type.  Lagrangians with this symmetry are called 
supersymmetric$^{8)}$.  Local (i.e., space-time dependent) supersymmetry 
necessarily involves gravity leading to supergravity theories$^{9)}$.  There 
are also continuous Lie symmetries which can be either local or global; 
the former case yields Yang-Mills gauge theories.  It is thought that the 
Lagrangian which describes all the fundamental interactions, save perhaps 
gravity, is \underbar {locally} invariant under a yet to be discovered Lie 
group.  The success of continuous Lie groups is linked to the existence of 
additively conserved quantum numbers.  Finally, discrete symmetries, 
although a definite logical possibility, have not found wide usage because 
they give rise to multiplicatively conserved quantum numbers, even though 
they have distinct theoretical advantages$^{10)}$.

The kinetic part of ${\cal L},~{\cal L}_{\rm{kin}}$, is the most symmetric 
as its symmetry is always greater (or equal if ${\cal L} = {\cal 
L}_{\rm{kin}}$) than that of ${\cal L}$.  Given $N$ Weyl spinors, the 
fermion kinetic term has a global $U(N)$ symmetry which includes chiral 
symmetry.  Not all of this symmetry can be gauged for fear of introducing 
unrenormalizable anomalies$^{11)}$.  The spinless boson kinetic term for 
$M$ bosons displays an $SO(M)$ global symmetry which can all be gauged 
(except for $M=6$).  It can happen that the number of spin-0 fields is 
twice the number of two-component Weyl spinors.  Then the symmetry is 
$0(M)\times$ global $N=1$ supersymmetry.  Thus ${\cal L}_{\rm{kin}}$ has 
an enormous symmetry which will be nibbled away by the other terms.  So it 
is convenient to discuss these terms by their action on the symmetries of 
${\cal L}_{\rm{kin}}$.

The gauge interaction terms, ${\cal L}_{\rm{int}}^{\rm g}$, of dimension-4, 
respects the chiral symmetry of ${\cal L}_{\rm{kin}}$ but will in general 
explicitly break the global supersymmetry of ${\cal L}_{\rm{kin}}$ (if 
any), which is then restored as the gauge coupling vanishes.

The Yukawa term, ${\cal L}_{\rm{int}}^Y$, explicitly breaks the chiral 
symmetries of ${\cal L}_{\rm{kin}}$ and in general the supersymmetry.  
This term is the most intriguing since the observed patterns of fermion 
masses mixing angles and CP violation$^{12)}$, depend on it.  Unlike gauge 
couplings, there is no known principle governing its form.  The 
restrictions imposed on it by the known gauge couplings unfortunately does 
not suffice to make it predictive.  This is where additional symmetries, 
such as the family symmetry will prove invaluable.  Certain patterns can 
be inferred from experiment:  the existence of low mass fermions indicates 
that chiral symmetry may not be too badly violated while the absence of 
low mass spinless particles suggests that supersymmetry is badly broken.

Finally, the most unprincipled part of ${\cal L}$ is that which describes 
the interactions among the spinless bosons.  It can occur with dimensions 
-4, -3, -2, and -1.  It can be used to achieve two types of symmetry 
breaking.  a)  Explicit breaking which is allowed for all except the gauge 
symmetries.  Two cases arise -- if the breaking term has dimension -4, 
quantum corrections will spread the breaking action into other parts of 
${\cal L}$.  This is the so-called ``hard breaking''.   When the breaking 
is done by terms of lower dimensions, (``soft breaking'') the quantum 
corrections to the relations implied by unbroken symmetry will be 
calculable$^{13)}$.  b) Spontaneous breaking.  In this case, the field 
equations are not affected, just the choice of ground state.  Its 
importance lies in the Higgs mechanism$^{14)}$ -- the only known way to 
break symmetry without losing renormalizability.  Spontaneous breaking can 
be used on all symmetries with one important proviso -- massless Goldstone 
bosons$^{15)}$ will appear if the symmetry is continuous and global.  Even 
then, there is an ingenious evasion mechanism:  match the global symmetry 
with a minor local symmetry and break both spontaneously, leaving the sum 
invariant.  In this way the Goldstone danger is avoided, but one is left 
with a global symmetry.  Such a mechanism is at work in the SU$_5$ model 
where due to the reducibility of the fermion representation, a global 
$U_1$ symmetry exists.  It is broken spontaneously together with a $U_1$ 
from within the $SU_5$, leaving a linear combination unbroken, which is 
baryon number minus lepton number.  Finally, let us mention that a ${\cal 
L}_{\rm{int}}^{\rm sc}$ made up of dimension -4 terms alone will give rise to 
spontaneous breaking of symmetry$^{16)}$, via quantum effects.  Discrete 
symmetries can be spontaneously broken without an ensuing Goldstone boson, 
and therein lies their theoretical attractiveness.  This breakup of ${\cal 
L}$ makes it convenient to discuss pseudo-symmetries$^{17)}$, i.e., 
symmetries respected by some parts of ${\cal L}$ but not by all.  For 
instance when the symmetry or ``${\cal L}_{\rm{kin}}$'' + ${\cal 
L}_{\rm{int}}^{\rm sc}$ is larger than that of ${\cal L}_{\rm{int}}^{\rm g}$ (where 
``${\cal L}_{\rm{kin}}$'' does not contain the vector self-interactions).  
The spontaneous breakdown via ${\cal L}_{\rm{int}}^{\rm sc}$ leads to more 
Goldstone bosons but those not eaten by the gauge fields acquire a mass 
due to the explicit breaking in ${\cal L}_{\rm {int}}^g$.  These are the 
pseudo-Goldstone bosons.  One might consider an analogous case with 
supersymmetry where ${\cal L} - {\cal L}_{\rm{int}}^{\rm g}$ is supersymmetric.  
Then one will obtain in this way, by turning on ${\cal L}_{\rm{int}}^{\rm g}$ 
``pseudo-Goldstinos''.  Lastly the soft explicit breaking occurs when 
${\cal L} = {\cal L}_0 + \tilde{\cal L}$ with $\tilde{\cal L}$ breaking 
the symmetry of ${\cal L}$ with fermion and spinless mass terms and 
possibly by cubic scalar self-interaction terms.

\noindent  II)  \underbar{The Zoo of Elementary Fermions - ${SU_5 \subset SO_{10} \subset E_6 \subset}$ ...}

According to $SU_3^{\rm c}$ ($c$ is for color), fermions come in two genres -- 
leptons with no color (color singlets) and quarks (antiquarks) which are 
color triplets (antitriplets).  No fermions with other color assignments 
are known at present, but their absence at low mass may just be a 
``caprice'' of the mass matrix and may not have any fundamental 
significance.  The masses of the observed fermions come in a rough pattern 
which enables us to define the concept of a family.  First we have very 
low mass fermions which under $SU_2 \times U_1 \times SU_3^{\rm c}$ transform as
\vfill\eject

$$({\sq2,~\sq1}^c)~~~~~({\sq 1,~\sq\overline 3}^c)~~~~~({\sq 1,~\sq\overline 
3}^c)~~~~~({\sq 2,~\sq3}^c)~~~~~({\sq 1,~\sq1}^c)$$
$$\left(\nu_{eL}\atop e_L\right)~~~~~~~~\overline {\sq 
d}_L~~~~~~~~~~~\overline {\sq u}_L~~~~~~~\left({\sq u}_L\atop {\sq 
d}_L\right)~~~~~~~~\overline e_L$$
[Notation: $e_L = 2$cpt-left-handed electron field, $\overline{e}_L = 
\sigma_2e_R^\ast =$ right-handed positron field; quark fields are 
underlined by a wiggle, to indicate color.]  We call this array the 
electron family.

The remarkable thing is that this pattern is repeated at a slightly higher 
mass, yielding the muon family with the same quantum numbers
$$\left(\nu_{\nu L}\atop \nu_L\right)~~~~~~\overline {\sq 
s}_L~~~~~~\overline{\sq c}_L~~~~~~\left({\sq c}_L\atop {\sq 
s}_L\right)~~~~~~\overline {\nu}_L\ .$$
Not shown here are the slight Cabibbo mixings of ${\sq d}_L$ and ${\sq 
s}_L$.  This family incorporates the GIM mechanism$^{18)}$.  As if it were 
not enough, it seems a third family is being discovered with a much higher 
central mass -- the $\tau$ family:
$$\left(\nu_{\tau L}\atop \tau_L\right)~~~~~~\overline {\sq 
b}_L~~~~~~\overline {\sq t}_L~~~~~~\left( {\sq t}_L\atop {\sq 
b}_L\right)~~~~~~\overline {\tau}_L\ ,$$
where the yet undiscovered charge 2/3 t-quark is heavier than 7 GeV (in 
the sense that no $t\overline {t}$ state exists at a mass of up to 15 
GeV).  Theoretically, the discovery of the t-quark at a reasonable mass 
would validate this emerging family picture.  Alternatively, its 
non-existence at a mass $\ltwid$ 30 GeV would give credence to the 
``topless'' models advocated by exceptional group enthusiasts$^{7)}$.  To 
conclude this preliminary classification, we note that neutrinos are all 
light although their families' central mass increases.  This could be due 
to the absence of right-handed partners so that the only way they can 
acquire mass is by developing a Majorana-type mass which has different 
quantum numbers.

Faced with these three families of fermions we now briefly review various 
attempts at defining the families themselves with unifying groups.  Then 
we will approach the problem of interaction between the families.

Since QCD and QFD are described by Yang-Mills theories, it is natural to 
consider a larger Yang-Mills theory which contains these two, i.e., the 
larger group of local invariance, G, will include $SU_2 \times U_1 \times 
SU_3^{\rm c}$ as a subgroup.  Let us for instance consider the imbedding of 
$SU_3^c$ with $G$ such that at least one representation of $G$ exists with 
at most $1^c,~3^c$ and/or $\overline{3}^c$, in order to represent the 
fermions.  These have been listed$^{19)}$, but only few are noteworthy:  
only in three cases do the quarks and leptons share weak interactions as a 
result of the group structure.  In all other cases, the quark-lepton 
universality of the weak interactions must arise from the specifics of the 
breaking mechanism of $G$ down to $SU_3^c \times U_1 \times SU_2$.  The 
three cases of interest are
$$G = SU_n \supset SU_{n-3} \times U_1 \times SU_3^c\ ,$$
with the fermions in the $({\sq n}x...x{\sq n})_A$ of $SU_n$.  Call it 
case I.  Then we have case II.
$$G = SO_n \supset SO_{n-6} \times U_1 \times SU_3^c\ ,$$
with the fermions in the spinor representation of $SO_n$.  For both cases, 
the electric charge ratio between quarks and leptons is arranged by hand 
through the use of the $U_1$ factor in the flavor group.  Case III 
includes the exceptional groups.  The relevant ones are
$$G = E_6 \supset SU_3 \times SU_3 \times SU_3^c\ ,$$
with the fermions in the (complex) ${\sq {27}}$, and
$$G = E_7 \supset SU_6 \times SU_3^c\ ,$$
with the fermions in the (pseudoreal) ${\sq 56}$.  In both cases, the factor 
of 3 between lepton and quark charges arises as a result of the group 
structure.  In this sense, these are the most natural simple Lie groups.

There is in the literature an example of each of the above three cases.  
The most studied$^{20)}$ and apparently successful is the $SU_5$ model of 
Georgi and Glashow, which is an example of case I.  The imbedding
$$SU_5\supset SU_2 \times U_1 \times SU_3^c$$
is defined by the fundamental of $SU_5$,
$${\sq 5} = ({\sq 2,~\sq1}^c) + ({\sq 1,\sq3}^c)\ ,$$
so that each family is described by a $\overline {\sq 5}$ and a ${\sq 10}$ 
(see Table I)
$${\sq 10} = ({\sq 5}\times{\sq 5})_A = ({\sq 1},\overline {\sq 3}^c) + 
({\sq 2,\sq3}^c) + ({\sq 1,\sq1}^c)\ .$$
This pattern is repeated thrice, once for each family.  The spin-1 bosons 
belong to the adjoint representation, which is
$${\sq 24} = ({\sq 1,\sq8}^c) + ({\sq 3,\sq1}^c) + ({\sq1,\sq1}^c) + ({\sq 2,\sq3}^c) 
+ ({\sq 2,}\overline{\sq3}^c)\ ,$$
corresponding to the gluon octet, the vectors of the GWSW model, and six 
others which play the dual role of causing lepton-quark and quark-quark 
transitions.  These cause proton decay in second order in the $SU_5$ 
coupling constant.  The fermion mass matrix consists of two parts, 
$\overline {\sq 5} \times {\sq 10} = {\sq 5} + 
\overline {\sq 45}$ which 
gives mass to the charged leptons and charge-1/3 quark within each family, 
and $({\sq 10}\times {\sq 10})_S = \overline {\sq 5} + 
{\sq 50}$ which 
gives a mass to the charge 2/3 quarks.  The minimal Higgs structure is a 
${\sq 24}$ to break $SU_5$ down to $SU_2\times U_1\times SU_3^c$ and a ${\sq 
5}$ to break $SU_2\times U_1$ down to $U_1^\gamma$.  Some consequences of 
this model are (at some scale)
$${m_d\over m_e} = {m_s\over m_\mu} = {m_b\over m_\tau} = 1\ ,~~~~~~~~{\rm 
sin}^2\theta_{\rm w} = {3\over 8}\ ,$$
where $\theta_{\rm w}$ is the Weinberg angle.  The theory is asymptotically free 
so that perturbation theory can be used reliably over large scales.  By 
matching the strong and weak coupling constants at our scale as a boundary 
condition, one finds$^{21)}$ that $SU_5$ symmetries are valid at very large 
masses of $0(10^{14}~{\rm GeV})^{22)}$ (i.e., very short distances).  Then one 
finds the proton decays with a rate of $10^{-32}$ per year for all modes.  
Similarly one finds$^{20)}$ renormalized values for sin$^2\theta_w \sim 
0.20 - .21,~{m_b\over m_\tau} \sim 2-3,~ {m_s\over m_\mu} \sim 4-5$.  
While these results are spectacular, the ratio ${m_d\over m_e}$ comes out 
all wrong, presumably because we are dealing with very light particles.  
As a further consequence of this theory with two ${\sq 5}$-Higgs 
fields$^{23)}$ one can explain the observed overabundance of matter over 
antimatter in the universe, while starting from symmetric boundary 
conditions.  Lastly, we re-emphasize an important aspect of the $SU_5$ 
theory:  the neutrinos are, as in the GWSW model, strictly massless 
because they are forbidden from acquiring a Majorana mass by the exact 
conservation of baryon number minus lepton number, $B-L$.  This law comes 
about because of the reducibility of the fermion representations which 
allows for a conserved quantum number which is 1 for the ${\sq 10}$, -3 for 
the $\overline {\sq 5}$ of fermions and -2 for the ${\sq 5}$ of Higgs.  When 
the ${\sq 5}$ of Higgs acquires a vacuum expectation value, this $U_1$ as 
well as the $U_1$ within $SU_5$ which has value 1 for the $(1,3^c)$ and 
-3/2 for the (2,1) of ${\sq5}$, are broken, but as the ${\sq 5}$ only has one 
non-zero entry, a linear combination is preserved:  $B-L$ for the 
fermions.  In a more unified theory where the fermion family not be 
reducible, this conservation law will not exist.  Then the neutrinos will 
be free to acquire Majorana masses.

The next group in this family description is $SO_{10}^{(5,6)}$ which falls 
in case II.  Unlike $SU_5$ it is automatically free of anomalies$^{5)}$.  The 
imbedding is given by
$$SO_{10} \supset SU_5 \times U_1\ ,$$
with the fermions appearing in the (complex) spinor representation (see 
Table II)
$${\sq 16} = \overline{\sq 5} + {\sq 10} + {\sq 1}\ .$$
Here, there is an extra neutral lepton helicity for each family.  It can 
act as a right-handed neutrino, which means that the neutrinos, in this 
theory, are not automatically massless.  To see this, consider the fermion 
mass matrix
$${\sq 16} \times {\sq 16} = ({\sq 10} + {\sq 126})_S + {\sq 120}_A\  .$$
The Higgs structure is very rich.  With just ${\sq 10}$'s of Higgs, the 
neutrino mass occurs in the same way as that of the charge 2/3 quark.  
Special measures have to be taken to insure the low mass of $\nu_L$ (these 
are the true leptons!).  One way$^{24)}$ is to use the Majorana mass of 
this extra right-handed neutrino, which appears in the $\sq{  126}$.  Call 
this extra lepton helicity $\overline{\nu}_L$.  Then the neutral lepton 
mass matrix is
$$\left( \nu_L^T~ \overline\nu_L^T\right)~~ \left( \matrix{0&a\cr
a&A\cr}\right)~~~\left( \nu_L\atop \overline {\nu}_L\right)\ ,$$
with $a$ proportional to $m_{2/3}$.  Take $a << A$.  Then $\nu_L$ is 
approximately massless and very slightly mixed.  This is, however, 
slightly disturbed by radiative corrections.  Note that a high scale is 
introduced in the mass matrix.  Another way$^{6)}$ is to invent yet another 
neutral lepton helicity which will act as the true Dirac partner of 
$\overline\nu_L$, which we label $Y_L$.  Then there are three neutral 
lepton helicities per family giving a mass matrix (per family),
$$\left( \nu_L^T~{\overline\nu}_L^T~Y_L^T\right)_{\left( \matrix{0&a&0\cr 
a&0&A\cr
0&A&0\cr}\right)~~\left(\matrix{\nu_L\cr  {\overline\nu}_L\cr   
Y_L\cr}\right)\ ,}$$
with again $a << A$.  The $\nu_L$ is slightly mixed but strictly massless. 
 This extra Diract mass introduces a 16 Higgs which is ten used to break 
$SO_{10}$ down to $SU_5$.  For a particular choice of ${\cal L}_{\rm 
int}^{sc}$, the fermion reducibility can again be used to produce a $B-L$ 
conservatin law, thereby forbidding Majorana masses.  In this scheme, the 
fermion mass matrix has an Abelian global family symmetry, $U_1\times 
U_1\times U_1$, which allows for a $t$-quark mass of the order of 15 GeV.

Finally, the most promising candidate of type III is based on $E_6^{(7)}$, 
with the imbedding
$$E_6 \supset SO_{10}\times U_1\ ,$$
with the fermions in the (complex) ${\sq 27}$ (see Table III)
$${\sq 27} = {\sq 16} + {\sq 10} + {\sq 1}\ .$$
This produces the singlet used in the preceeding model, but introduces for 
each family ten additional helicities.  Interestingly, a feature of the 
GWSW model that was lost in $SU_5$ and $SO_{10}$ is regained:  all the 
symmetry breaking needed can occur in the fermion-fermion operator with, 
for instance, ${\sq 16} \times {\sq 1}$ breaking $E_6$ down to $SU_5$, and 
the ${\sq 10}\times {\sq 10}$ which contains a ${\sq 24}$ of $SU_5$ breaking 
$SU_5$ down to $SU_2\times U_1\times SU_3^c$.  The Higgs structure of the 
model is found in
$${\sq 27}\times {\sq 27} = (\overline{\sq 27} + 
{\sq 351}')_s + {\sq351}_A\ .$$
Spinless bosons transforming as the ${\sq 27}$ are not sufficient to give the 
particles their requisite masses.  At least one ${\sq 351}'$ of Higgs is 
needed to get ${\sq 16}\times {\sq 1}$ and ${\sq 10}\times 
{\sq 10}$ 
terms.  Then, with these families, the theory is just asymptotically free. 
 Addition of another fermion family, or of another ${\sq 351}'$ changes the 
sign of the derivative of the running gauge coupling constant.  The 
consequences of a temporarily free theory are not understood, but one may 
speculate that the attrative and repulsive vector forces may compete with 
that of gravity, leading to the ``bounces'', thereby at least delaying the 
formation of singularities.  Although much work has been done on the mass 
matrix of the $E_6$ model, there is no obvious way to reproduce the 
neutral lepton mass matrix.  Still one should not be discouraged as it 
presents the greatest unification of the fermions.  However, none of these 
attempts provides a reason for the triplication of families.

\noindent  III)  \underbar{The search for family symmetries}

After having presented certain ``unified'' descriptions of the fermions, 
we are still not quite unified enough because of the apparent triplication 
(and possibly infinite xeroxing) of families.  As the concept of families 
arises by looking at the fermion masses, we must search for ways to narrow 
down ${\cal L}_{\rm int}^Y$.  This term is the most arbitrary in the 
$SU_2\times U_1$ model in which an enormous number of parameters is 
introduced.  Several attempts have been made to introduce a family 
symmetry, both discrete$^{10}$ and continous.  In one of these$^{25)}$, a 
gauged family symmetry (called horizontal symmetry) based on $S0_3$ is 
introduced at the $SU_2\times U_1\times SU_3^c$ level.  However, this 
scheme, because of the symmetries of the fermion mass matrix cannot be 
extended directly to $SU_5$ (unless one wants to introduce three ${\sq 
50}'$s of Higgs!).  Another, previously mentioned$^{6)}$, uses a global 
$U_1\times U_1\times U_1$ family symmetry in $SO_{10}$.

Although it is not clear at what stage one should introduce the family 
symmetry, we find it convenient to do it at the level of $SU_5$ because of 
the regularities in the mass matrix, where we have just the nearly 
massless $e$-family and the heavier $\mu$- and $\tau$-families.  We 
propose to use the family group to tell us why the $e$-family is so light 
and at the same time reduce the number of parameters in ${\cal L}_{\rm 
int}^Y$.  The family group may or may not be gauged.  At least it should 
be a symmetry of the terms of dimension -4 in ${\cal L}$.  If it is 
gauged, we should beware of anomalies since the family spinors are Weyl 
spinors.  Also it must be badly broken to avoid at low energy flavor 
changing forces.  This last aspect is neatly done by having the agent that 
does the superstrong breaking do at the same time the family breaking.  We 
now present an illustrative example of this type.

Let $SU_2$ be the family symmetry$^{26)}$.  Assume that the two lightest 
$SU_5$ families form a doublet under it.  The fermion content in terms of 
$SU_2^f\times SU_5$ is then
$$\eqalign{ F \equiv &\left( {F_e\atop F_\mu}
\right) \sim ({\sq 2}, {\sq10})~~;~~f \equiv \left( {f_e\atop 
f_\mu}\right) \sim ({\sq 2}, \overline{\sq 5})\cr
& F_\tau \sim ({\sq 1}, {\sq 10})~~~~~~~~~~f_\tau \sim ({\sq 1}, 
\overline{\sq 5})\ .\cr}$$
The spinless bosons are taken to be
$$\eqalign{h \sim ({\sq 1}, {\sq 5})~~~~~~~~~~&h'\sim ({\sq 1}, {\sq 5})\cr
H\sim ({\sq 2}, {\sq 5})~~~~~~~~~~&H' \sim ({\sq 2}, {\sq 5})\ .\cr}$$
as well as $\Phi \sim ({\sq 2,} {\sq 24})$.  $\Phi$ acquires a very large 
vacuum expectation value, breaking $SU_5$ down to $SU_2\times U_1\times 
SU_3^c$ and breaking the family $SU_2$.  The Yukawa interactions are given 
by the terms $F_\tau F_\tau h,~F_\tau f_\tau {\overline h}^\prime,~F_\tau 
FH,~ F_\tau f{\overline H}^\prime,~Ff_\tau {\overline H}^\prime$, which 
besides $SU_2^f\times SU_5$, respect three global $U_1$'s.  One of those 
is explicitly broken in ${\cal L}_{\rm int}^{sc}$ by a term of the form 
${\overline h}h^\prime{\overline H}H^\prime$; the other two, $X$ and $Y$ 
are summarized in the table
$$\matrix{&h&g^\prime&H&H^\prime&F_\tau&f_\tau&F&f\cr
X~~~~&1&1&1&1&-1/2&3/2&-1/2&3/2\cr
Y~~~~&1&-1&-1&1&-1/2&-1/2&3/2&3/2\cr}$$
Both $X$ and $Y$ are to be broken spontaneously.  The breaking of $X$ does 
not give rise to a Goldstone boson because $X$-conservation is replaced by 
$B-L$ conservation.  $Y$-breaking gives rise to a Goldstone boson which 
acquires mass by QCD instanton effects$^{27)}$.  In view of the 
non-existence of a low mass axion$^{28)}$ it might be necessary to add a 
Higgs mass of the form $h\overline h^\prime$ which explicitly breaks $Y$ 
(this can be used to obtain the desired sign for the $u-d$ mass 
difference).  One interesting consequence of this model is that the 
$e$-family is massless in lowest order but acquires calculable (e.g., 
finite) corrections, by means of, among others, terms of the form 
$FFH{\overline H}^\prime h^\prime$, $FFH^2h$, $FfH{\overline H}^\prime 
{\overline h}$, $Ff{\overline H}^{\prime 2} h$, consistent with all the 
symmetries of ${\cal L}$.  Such a model alters the value of $m_e/m_d$ 
without affecting the others.  Also, it yields small $u$ and $d$ masses, 
accounting for isospin symmetry.  This $SU_2$ family symmetry looks more 
natural in an $SO_{10}$ model for there only the $SU_2$ forbids the 
$e$-family from acquiring a mass (whereas here $Y$ is needed) by having 
Higgs with $({\sq 1},{\sq 10})$ and $({\sq 2},{\sq 10})$ only.  This model is 
presented here as an illustration of the uses one may make of family 
symmetries.  Note that a gauged $SU_3$ family symmetry is feasible 
provided one uses an anomaly-free set of representations.  Finally, one 
may unify the family symmetry with the $SU_5$.  There is a strong 
candidate is $SU_8$, with the fermions appearing in 
${\sq 56}_L + \overline {\sq 56}_L^\prime.$  We hope to present these 
models in detail in a later publication.

In closing, let us mention what we have not discussed:  the all-important 
problem of hierarchies$^{29)}$ for which this survey offers no answers nor 
provides any clues.

\noindent \underbar  {Acknowledgement}

The author wishes to thank his collaborators, Professors M. Gell-Mann and 
R. Slansky for invaluable comments, suggestions and ideas concerning the 
material presented here.
\vfill\eject

\centerline { Table I}
\centerline {\underbar {$SU_5$ Unification Picture}}
\bigskip
\noindent $SU_5 \supset SU_2 \times U_1 \times SU_3^c$

$$~~~~~\left(\nu^{}_{eL}\atop e^{}_L\right)~~~~~~~~~\overline {\sq 
d}_L~~~~~~~~~~~~~~~~~~\overline {\sq u}_L~~~~~~~\left({\sq u}_L\atop {\sq 
d}_L\right)~~~~~~~~\overline e_L$$
$$~~~~~~~~\underbrace{(\sq 2,~\sq 1^c)~~+~~(\sq 1,~\sq\overline 3^c)}
~~~~~~~~~~~~~\underbrace{({\sq 1,~\sq\overline 
3}^c)~~+~~({\sq 2,~\sq 3}^c)~~+~~({\sq 1,~\sq1}^c)}$$
$$\overline{\sq 5}~~~~~~~~~~~~~~~\hskip 2cm~~~~~~~~~~~~~~~~\sq{10}$$

\vskip 1cm
$$~~~~~\underbrace{\left(\nu^{}_{\mu L}\atop \mu^{}_L\right)~~~~~~~~~\overline {\sq 
s}_L}~~~~~~~~~~~~~~~~~~\underbrace{\overline {\sq c}_L~~~~~~~\left({\sq c}_L\atop {\sq 
s}_L\right)~~~~~~~~\overline \mu_L}$$
$$~\overline{\sq 5}~~~~~~~~~~~~~~\hskip 2cm~~~~~~~~~~~~~~~~\sq{10}$$

\vskip 1cm
$$~~~~~\underbrace{\left(\nu^{}_{\tau L}\atop \tau^{}_L\right)~~~~~~~~~\overline {\sq 
b}_L}~~~~~~~~~~~~~~~~~~\underbrace{\overline {\sq t}_L~~~~~~~\left({\sq t}_L\atop {\sq 
b}_L\right)~~~~~~~~\overline \tau_L}$$
$$~~\overline{\sq 5}~~~~~~~~~~~~~\hskip 2cm~~~~~~~~~~~~~~~~\sq{10}$$

\vfill\eject
\centerline { Table II}
\centerline { \underbar {SO}$_{10}$ \underbar{Unification}}
\bigskip
\noindent $SO_{10} \supset SU_5\times U_1$

$$~~~\underbrace{\left(\nu^{}_{eL}\atop e^{}_L\right)~~~~~~~~~\overline {\sq 
d}_L}~~~~~~~~~~\underbrace{\overline {\sq u}_L~~~~~~~\left({\sq u}_L\atop {\sq 
d}_L\right)~~~~~~~~\overline e_L}~~~~~~~~~\overline\nu_{eL}^{}$$
$$\hskip 1cm~~~~\overline{\sq 5}~~~~~~~~~~\hskip 2cm~~~~~~~~~~~\sq{10}~~~~~~~~~~\hskip 2cm~{\sq 1}$$
$$\underbrace{\hskip 12cm}$$
\centerline{$\sq 16$}
\vskip 1cm

$$~~~\underbrace{\left(\nu^{}_{\mu L}\atop \mu^{}_L\right)~~~~~~~~~\overline {\sq 
s}_L}~~~~~~~~~~\underbrace{\overline {\sq c}_L~~~~~~~\left({\sq c}_L\atop {\sq 
s}_L\right)~~~~~~~~\overline \mu_L}~~~~~~~~~\overline\nu_{\mu L}^{}$$
$$\underbrace{\hskip 12cm}$$
\centerline{$\sq 16$}
\vskip 1cm
$$~~~\underbrace{\left(\nu^{}_{\tau L}\atop \tau^{}_L\right)~~~~~~~~~\overline {\sq 
b}_L}~~~~~~~~~~\underbrace{\overline {\sq t}_L~~~~~~~\left({\sq t}_L\atop {\sq 
b}_L\right)~~~~~~~~\overline \tau_L}~~~~~~~~~\overline\nu_{\tau L}^{}$$
$$\underbrace{\hskip 12cm}$$
\centerline{$\sq 16$}

\vfill\eject

\centerline {Table III}
\centerline {\underbar {E}$_6$ \underbar{Unification}}
\bigskip
\noindent $E_6 \supset SO_{10} \times U_1$

$$~~~\underbrace{\left(\nu^{}_{eL}\atop e^{}_L\right)~~~\overline {\sq 
d}_L~~~~~~~\overline {\sq u}_L~~~\left({\sq u}_L\atop {\sq 
d}_L\right)~~~\overline e_L~~~~~~~\overline\nu_{eL}^{}}
~~~~\underbrace{Y_{eL}}~~~~\underbrace{\left(N^{}_{eL}\atop E^{-}_L\right)~~~\overline{\sq h}_L~~~
\left(\overline N^{}_{eL}\atop \overline N^{}_{eL}\right)~~~{\sq h}_L}$$
$$~~~~~~~~~~~~~~~~~~\underbrace{{\sq 16}\hskip 4.6cm{\sq 1}\hskip 3.5cm {\sq 10}}~~~~$$
\centerline{$~~~~~~~~~~~~~~~\sq{27}$}
\vskip 1cm

$$~~~\underbrace{\left(\nu^{}_{\mu L}\atop \mu^{}_L\right)~~~\overline {\sq 
s}_L~~~~~~~\overline {\sq c}_L~~~\left({\sq c}_L\atop {\sq 
s}_L\right)~~~\overline \mu_L~~~~~~~\overline\nu_{\mu L}^{}}
~~~~\underbrace{Y_{\mu L}}~~~~\underbrace{\left(N^{}_{\mu L}\atop M^{-}_L\right)~~~\overline{\sq k}_L~~~
\left(\overline M^{}_{eL}\atop \overline N^{}_{\mu L}\right)~~~{\sq k}_L}$$
$$~~~~~~~~~~~~~~~~~~\underbrace{{\sq 16}\hskip 4.6cm{\sq 1}\hskip 3.5cm {\sq 10}}~~~~$$
\centerline{$~~~~~~~~~~~~~~~\sq{27}$}
\vskip 1cm

$$~~~\underbrace{\left(\nu^{}_{\tau L}\atop \tau^{}_L\right)~~~\overline {\sq 
b}_L~~~~~~~\overline {\sq t}_L~~~\left({\sq t}_L\atop {\sq 
b}_L\right)~~~\overline \tau_L~~~~~~~\overline\nu_{\tau L}^{}}
~~~~\underbrace{Y_{\tau L}}~~~~\underbrace{\left(N^{}_{\tau L}\atop T^{-}_L\right)~~~\overline{\sq j}_L~~~
\left(\overline T^{}_{\tau L}\atop \overline N^{}_{\tau L}\right)~~~{\sq j}_L}$$
$$~~~~~~~~~~~~~~~~~~\underbrace{{\sq 16}\hskip 4.6cm{\sq 1}\hskip 3.5cm {\sq 10}}~~~~$$
\centerline{$~~~~~~~~~~~~~~~\sq{27}$}
\vskip 1cm

\vfill\eject

\centerline {\underbar{ References}}

{\obeylines

\noindent{1.} J. C. Pati and A. Salam, Phys. Rev. ${\underline {D8}}$ (1973) 1240;$\underline{D10}$(1974) 275.
 
\noindent{2.} H. Georgi and S. L. Glashow, Phys. Rev. Lett. ${\underline {32}}$ (1974) 438.
 
\noindent{3.} Y. Nambu in ``Preludes in Theoretical Physics,'' ed. A. de Shalit (North-Holland, Amsterdam, 1966); 
H. Fritzsch and M. Gell-Mann, Proc. of the XVI International Conference on High Energy Physics, Vol. 2, p. 35 (national Accelerator Laboratory).
 
\noindent{4.} S. L. Glashow, Ph.D. Thesis, Harvard University 1959, and Nucl. Phys. ${\underline {22}}$ (1961) 579; 
S. Weinberg, Phys. Rev. Lett. ${\underline {19}}$ (1967) 1264; 
A. Salam, Proc. 8th Nobel Symp., Stockholm, ed. N. Svartholm 
(Almquist and Wiksells, Stockholm 1968) p. 367; A. Salam and J. C. Ward, 
Phys. Lett. ${\underline {13}}$ (1964) 168.
 
\noindent{5.} H. Fritzsch and P. Minkowski, Ann. of Phys. ${\underline {93}}$ (1975) 193; 
H. Georgi, Particles and Fields, 1974 (APS/DPF Williamsburg) ed. C. 
E. Carlson (AIP New York, 1975) p. 575; 
M. S. Chanowitz, J. Ellis and M. K. Gaillard, Nucl. Phys. ${\underline {B128}}$ (1977) 506.
 
\noindent{6.} H. Georgi and D. V. Nanopoulos, Harvard preprints 1978-1979.
 
\noindent{7.} F. G\"ursey, P. Ramond and P. Sikivie, Phys. Lett. ${\underline {B60}}$ (1975) 177; 
F. G\"ursey and M. Serdaro\v glu, Yale preprint, 1978; 
Y. Achiman and B. Stech, Phys. Lett. ${\underline {77B}}$ (1978) 389;
 Q. Shafi, Univ. of Freiburg preprint 1978.
 
\noindent{8.} For a review see P. Fayet, S. Ferrara, Phys. Reports ${\underline {32C}}$ (1977) 249.
 
\noindent{9.} D. Z. Freedman, P. van Nieuwenhuizen, and S. Ferrara, Phys. Rev. ${\underline {D13}} $(1976) 3214; 
S. Deser and B. Zumino, Phys. Lett. ${\underline {B62}}$ (1976) 335.
 
\noindent{10.} S. Pakvasa and H. Sugawara, Phys Lett. ${\underline {73B}}$ (1978) 61, 
Wisconsin preprint COO-881-66, 1978; 
H. Sato, University of Tokyo preprint, 1978; 
E. Derman, Rockefeller preprint, 1978.
 
\noindent{11.} H. Georgi and S. . Glashow, Phys. Rev. ${\underline {D6}}$ (1973) 429.
 
\noindent{12.} N. Kobayashi and K. Maskawa, Progr. Teor. Phys. ${\underline {49}}$ (1973) 652.
 
\noindent{13.} K. Symanzik in Coral Gables Conference on Fundamental Interactions at High Energies II.  (A. Perlmutter, G. J. Iverson and R. M. Williams eds., Gordon and Breach, 1970.)
 
\noindent{14.} P. W. Higgs, Phys. Rev. Lett. ${\underline {12}}$ (1964) 132; 
F. Englert and R. Brout, Phys. Rev. Lett. ${\underline {13}}$ (1964) 321; G. S. Guralnik, C. R. Hagen, and T. W. B. Kibble, Phys. Rev. Lett. ${\underline {13}}$ (1964) 585.
 
\noindent{15.} J. Goldstone, Nuovo Cimento ${\underline {19}}$ (1961) 154.
 
\noindent{16.} S. Coleman and E. Weinberg, Phys. Rev. ${\underline {D7}}$ (1973) 1888; 
E. Gildener and S. Weinberg, Phys. Rev. ${\underline {D13}}$ (1976) 3333.
 
\noindent{17.} S. Weinberg, Phys. Rev. Lett. ${\underline {29}}$ (1972) 1698.
 
\noindent{18.} S. L. Glashow, J. Illiopoulos, L. Maiani, Phys. ${\underline {D2}}$ (1970) 1285.
 
\noindent{19.} M. Gell-Mann, P. Ramond, and R. Slansky, Rev. Mod. Phys. ${\underline{50}}$ (1978)721.
 
\noindent{20.} A. J. Buras, J. Ellis, M. K. Gaillard, and D. V. Nanopoulos, Nucl. Phys. ${\underline {B135}}$ (1978) 66.
 
\noindent{21.} H. Georgi, H. R. Quinn, and S. Weinberg, Phys. Rev. Lett. ${\underline {33}}$ (1974) 451.
 
\noindent{22.} T. Goldman and D. Ross, Caltech preprint, CALT-68-704 (1979).
 
\noindent{23.} M. Yoshimura, Phys. Rev. Lett. ${\underline {39}}$ (1977) 1385; 
S. Dimopoulos and L. Susskind, SLAC-PUB-2126 (1978); 
D. Toussaint, S. B. Treiman, F. Wilczek and A. Zee, Princeton preprint 1978; 
J. Ellis, M. K. Gaillard and D. V. Nanopoulos, CERN preprint, TH-2596 (1978); 
S. Weinberg, Harvard preprint, HUTP-78/A040, 1978.
 
\noindent{24.} M. Gell-Mann, P. Ramond, and R. Slansky, unpublished.
 
\noindent{25.} F. Wilczek and A. Zee, Phys. Rev. Lett. ${\underline {42}}$(1979)421.
 
\noindent{26.} This concept has been used earlier under the name of M-spin.  
See for instance F. G\"ursey and G. Feinberg, Phys. Rev. ${\underline {128}}$ (1962) 378; 
T. D. Lee, Nuovo Cimento ${\underline {35}}$ (1965) 975; 
S. Meshkov and S. P. Rosen, Phys. Rev. Lett. ${\underline {29}}$ (1972) 1764; Phys. Rev. ${\underline {D10}}$ (1974) 3520.

\noindent{27.} G. 't Hooft, Phys. Rev. Lett. ${\underline {37}}$ (1976) 8.
 
\noindent{28.} R. D. Peccei and H. Quinn, Phys. Rev. Lett. ${\underline {38}}$ (1977) 40; Phys. Rev. ${\underline {D16}}$(1977)1791;
 S. Weinberg, Phys. Rev. Lett. ${\underline {40}}$ (1978) 22; 
F. Wilczek, Phys. Rev. Lett. ${\underline {40}}$ (1978) 279.

\noindent{29.} E. Gildener, Phys. ${\underline {D14}}$ (1976) 1667.
}

\end